\documentclass{article}

\usepackage{subcaption}
\usepackage[abs]{overpic}
\usepackage{color}

\title{High-speed hyperspectral four-wave-mixing microscopy with frequency combs}

\author{Brad C. Smith\textsuperscript{1,*}, Bachana Lomsadze\textsuperscript{1,2},\\ Steven T. Cundiff\textsuperscript{1,**}\\
\normalsize\textsuperscript{1} Department of Physics, University of Michigan,\\
\normalsize Ann Arbor, MI 48109, USA.\\
\normalsize\textsuperscript{2} Department of Physics, Santa Clara University,\\
\normalsize Santa Clara, CA 95053, USA.\\
\normalsize\textsuperscript{*}Current address: Lockheed Martin Coherent Technologies,\\
\normalsize 135 S Taylor Ave Louisville, CO 80027\\
\normalsize\textsuperscript{**}Corresponding author: cundiff@umich.edu
}

\begin{document}

\maketitle

\begin{abstract}
	A four-wave-mixing, frequency-comb-based, hyperspectral imaging technique that is spectrally precise, potentially rapid, and can in principle be applied to any material, is demonstrated in a near-diffraction-limited microscopy application.
\end{abstract}

\section{Introduction}
As the cost, robustness, and capability, of optical technologies continue to improve, their applications become increasingly widespread. In particular, the combination of spectroscopy and imaging, called hyperspectral imaging, is being applied to a diverse and growing number of fields including, but not limited to, agriculture \cite{Bauriegel2014}, food inspection \cite{Sun2010}, biology \cite{Vasefi2016}, and astronomy \cite{Garmire2003}. There are even commercial, hand-held hyperspectral imagers available now \cite{Specim2018} and detailed open-access instructions to build your own imager using 3D-printed and off-the-shelf parts \cite{Sigernes2018}.

All of these examples exploit linear optical phenomena (namely absorption and refraction), which make them simple but also limit their capabilities compared to nonlinear techniques. In the context of imaging, nonlinear methods have intrinsically higher spatial resolution, higher sensitivity to the environment, and the ability to probe richer information such as coupling between energy levels. For example, coherent anti-Stokes Raman spectroscopy (CARS) imaging \cite{Smith2012} probes the Raman response of the sample. There have also been hyperspectral images formed using multi-dimensional coherent spectroscopy (MDCS) \cite{Kasprzak2010}, which revealed coherent coupling between distant excitons, but the technique requires delay stages thereby limiting its throughput rate.

A consistent goal in all of these cases is to increase the speed, signal-to-noise ratio (SNR), and resolution (both spectral and spatial). All of these can be achieved by harnessing the bandwidth compression (i.e. the conversion of optical data with a large bandwidth -- commonly many THz -- to rf data with a significantly smaller bandwidth -- commonly tens to hundreds of MHz) of frequency combs in the experimental design \cite{Coddington2016}. An experiment similar to the CARS example mentioned above was able to do just this (Ref.\ \cite{Barlow2014}) and successfully achieved the same benefits of using combs as dual-comb spectroscopy did with respect to Fourier-transform infrared spectroscopy \cite{Coddington2016}. However, it also had the fundamental limitation of any CARS technique -- it relied on Raman shifts, which greatly limits the number of samples to which it can be applied.

Here, a four-wave mixing (FWM) based hyperspectral imaging technique is presented that can in principle be applied universally to any material while still retaining the potential speed, precision, and SNR advantages achievable with frequency combs.

\section{Experiment}
\label{sec:method}

One of the fundamental challenges for nonlinear optical spectroscopy lies in distinguishing a nonlinear signal from a linear one. CARS uses the anti-Stokes Raman shift to spectrally shift the FWM signal away from the pumps enabling the use of a simple optical filter to isolate the FWM signal for high sensitivity detection. In this work, we combine a comb-based version of ``frequency tagging'' \cite{Nardin2013} with the standard dual-comb read-out technique \cite{Coddington2016}. The technique is similar to that in Ref.\ \cite{Lomsadze2017}, but with the several modifications. 

\begin{figure}[ht]
\centering
\hspace*{\fill}
\begin{overpic}[scale=0.34, unit=1mm]{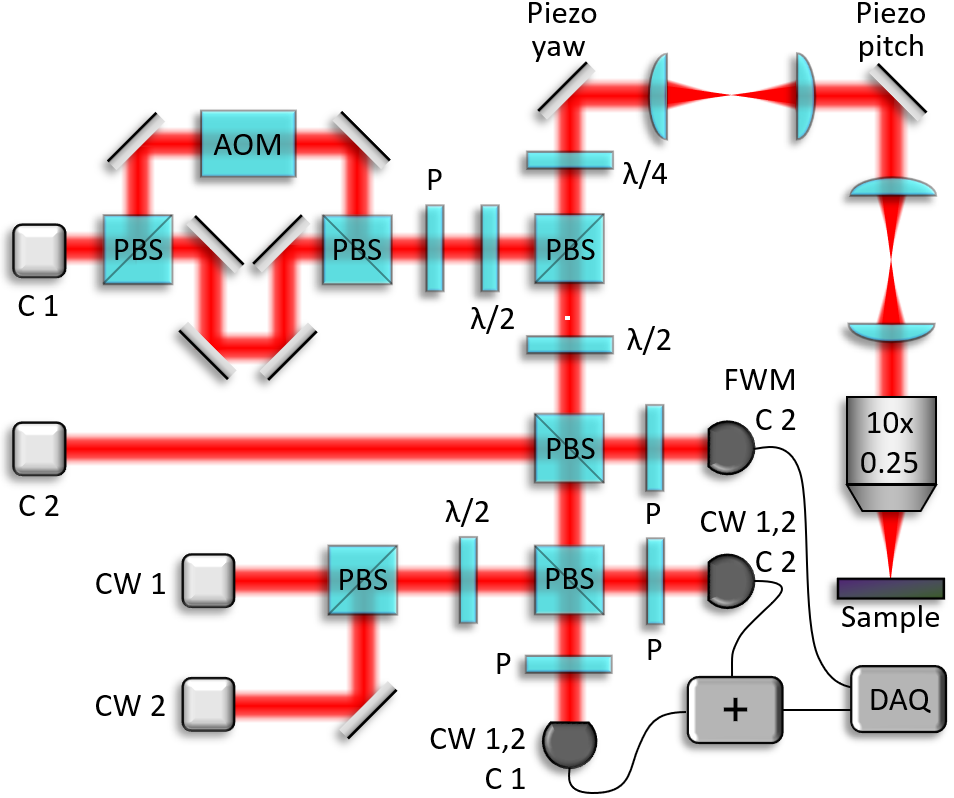}
 \put (0,68.5) {(a)}
\end{overpic}
\hspace*{\fill}

\vspace{5 mm}
\hspace*{\fill}
\begin{overpic}[scale=0.25, unit=1mm]{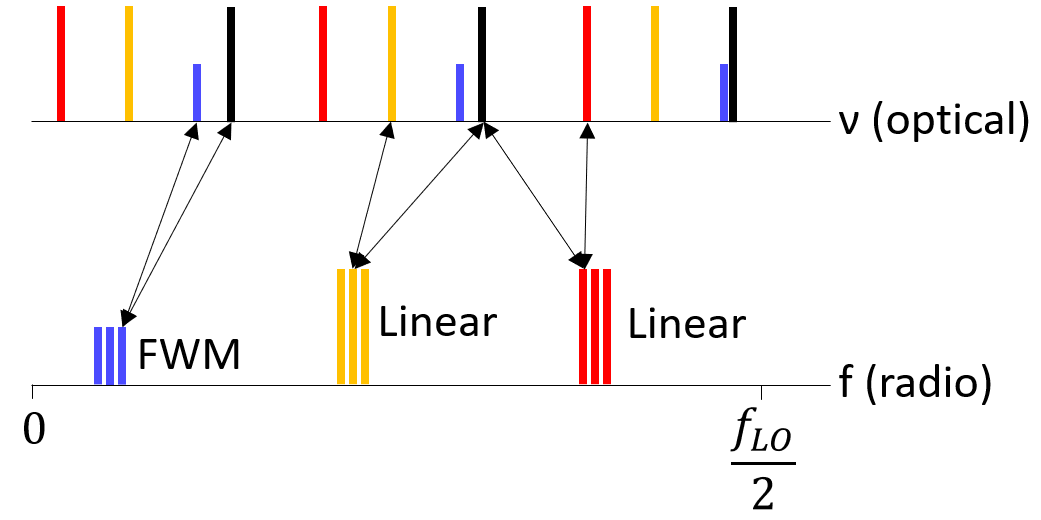}
 \put (-8,35) {(b)}
\end{overpic}
\hspace*{\fill}
\caption{(a) Simplified schematic diagram of experiment. C 1 = comb 1, C 2 = comb 2, PBS = polarizing beam splitter, AOM = acousto-optic modulator, P = polarizer, $\lambda/4$ = quarter-wave plate, and $\lambda/2$ = half-wave plate. Written above each detector are the combinations of optical signals of interest. A $10\times$ objective with a numerical aperture of 0.25 was used. (b) Mapping between optical comb teeth (red - comb 1, yellow - AOM shifted comb 1, blue - relevant FWM comb, black - comb 2) and the RF beat notes observed on a detector. Not shown is a large ``time-zero'' beat-note at 13 MHz (93 MHz repetition rate less the 80 MHz AOM frequency) corresponding to beating between the AOM-shifted and non-AOM-shifted pump comb lines.}
\label{fig:hyp1}
\end{figure}

In Ref.\ \cite{Lomsadze2017} two home-built Kerr-lens mode-locked Ti:Sapphire lasers with slightly different repetition rates were used. The repetition frequencies of the combs were phase locked to a direct digital synthesizer using feedback loops. The comb offset frequencies were not stabilized. The output of one of the combs was split into two parts using a half wave plate and a polarizing beam splitter (PBS). The offset frequency of the first part was shifted by an Acousto-Optical Modulator (AOM) and recombined with the second part on a PBS. The optical path lengths for the two arms were adjusted to overlap the two pulse trains in time. Before interacting with the sample, the beams were projected to the same linear polarization state using a polarizer. The combined beams were then focused on a sample consisting of 10 layers of 10 nm GaAs quantum wells (QW) separated by 10 nm thick Al$_{0.3}$Ga$_{0.7}$As barriers. The FWM signal that was emitted in the forward (phase-matched) direction and the incident beams were combined with another comb and interfered on a photodetector.  The down-converted linear and FWM signals were then spectrally separated in the RF domain (Fig.\ \ref{fig:hyp1}b). In the experiment, the relative phase noise between two combs caused by relative offset and repetition frequencies’ fluctuations were measured and corrected using a single continuous wave (CW) laser.

In the experiment described here, we made several modifications (see Fig.\ \ref{fig:hyp1}a). First, a second CW laser (Toptica DL100 external cavity diode laser) was used to greatly improve the usable spectral bandwidth through post-processing \cite{Deschenes2010}. Second, steering mirrors with piezoelectric actuators combined with two 4-f imaging systems enabled fast raster-scanning of the laser spot across the sample. Third, the FWM signal was detected in the backward direction making it compatible with standard imaging modalities (e.g.~microscopy). Note that although the FWM signal is phase-matched in the forward direction for this collinear excitation scheme, the strong absorption of the sample limits the effective excitation depth to an amount comparable to the wavelength of the pump light within the medium (approximately 200 nm). Thus, appreciable FWM can be emitted in the backward direction \cite{Honold1998}.

\begin{figure}
    \centering
    \begin{subfigure}[b]{\columnwidth}
        \includegraphics[scale=0.75]{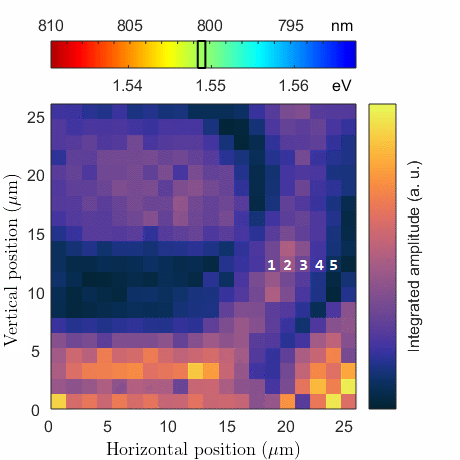}
    \end{subfigure}\\[\baselineskip]
    
    \centering
    \begin{minipage}[b]{\columnwidth}
      \begin{subfigure}[b]{0.5\columnwidth}
            \begin{overpic}[scale=0.5]{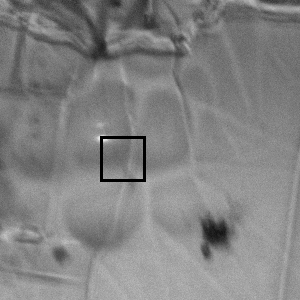}
             \put (98,1.5) {\color{white} 6 K}
            \end{overpic}
      \end{subfigure}
      \hfill
      \begin{subfigure}[b]{0.5\columnwidth}
            \begin{overpic}[scale=0.5]{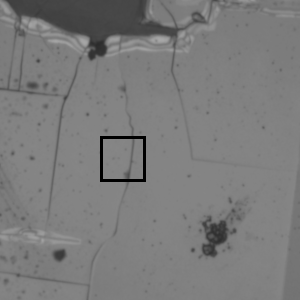}
             \put (90,1.5) {\color{white} 296 K}
            \end{overpic}
      \end{subfigure}
    \end{minipage}
    \caption{Top: 400-pixel hyperspectral image from Visualization 1 showing the spectrally integrated amplitude of the fully corrected FWM-LO heterodyne signal (6 K - in high-vacuum cryostat). 75 ms of data (6 bursts) were collected per pixel. Prior to integration, the spectrum was multiplied with a Gaussian whose full width at half maximum is represented by the black box on the upper colorbar. Normalized FWM spectra for 5 neighboring pixels (indicated by 1-5 in the image) are shown in Fig.~\ref{fig:FWM}. Bottom: Corresponding white-light images during experiment (6 K - in high-vacuum cryostat) and afterwards at room temperature (296 K - 1 atm pressure). The black square represents the hyperspectral scan area.}
    \label{fig:r0}
\end{figure}

\begin{figure}
    \centering
    \includegraphics[width=\columnwidth]{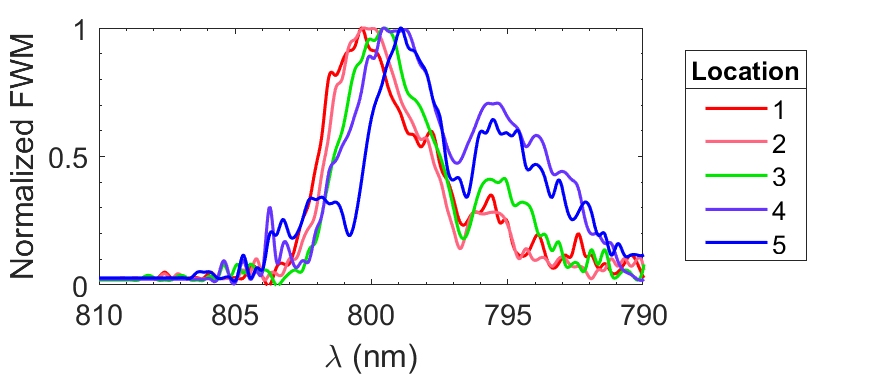}
    \caption{Normalized FWM spectra for five neighboring pixels who's positions are indicated in Fig.~\ref{fig:r0}}
    \label{fig:FWM}
\end{figure}

To validate the experimental technique, hyperspectral images were taken of a stack of 10 GaAs/AlGaAs quantum wells with 10 nm separation mounted on a sapphire substrate. The sample was cooled to <10 K in a high-vacuum, flow cryostat to enhance the spectral features of the $n=1$ light hole and heavy hole excitons \cite{Cundiff2008}. For more-interesting images, 6 different high-strain areas of the sample were selected. The pump and probe combs used were home-built Ti:sapphire oscillators centered around 800 nm with bandwidths of greater than 50 nm, output powers of about 100 mW, and loosely-locked repetition rates of about 93.5 MHz (differing by about 65 Hz). Two 24 nm optical band-pass filters centered around 800 nm were used -- one filtered the pump beam to avoid exciting unwanted transitions and the other filtered the FWM signal and local oscillator (LO) comb to reduce the amount of light sent to the detector. A wavemeter (Bristol) measured the optical frequencies of the CW lasers to provide coarse spectral calibration for the acquired data (center frequencies of 370.404(76) THz and 377.115(38) THz). A Labview program raster scanned the laser spot, waited several hundred ms after the movement command to allow settling of the physical parts, then acquired data from the data acquisition (DAQ) board, then repeated until 400 pixels of data were recorded. Due to the large size of the files that had to be saved after digitization for each pixel, each hyperspectral image took about 30 minutes of laboratory time to record even though only about 45 seconds worth of data was collected. This excessive acquisition time could be reduced using real-time techniques \cite{Roy2012,Ideguchi2014,Shen2018} that allow for significantly smaller file sizes. Further acquisition speed gains could come via faster raster scan control enabled for example by resonant galvanometer mirrors. The resulting data was post-processed similar to Ref.~\cite{Lomsadze2017}, with the added technique from Ref.~\cite{Deschenes2010}. Furthermore, due to computer memory limitations and processing speed considerations, much of the data between bursts in each FWM-LO heterodyne RF comb was ignored by splitting it up into 30 ps slices centered around each temporal burst (typically 6-10 slices total). Each slice was corrected separately, then they were all coherently combined by lining up the phases at each signal's peak. Lastly, a Gaussian window with a full width at half maximum (FWHM) of ~10 ps was applied to reject noise. This limited the spectral point spacing of the acquired hyperspectral images, but was still sufficient to resolve the light hole and heavy hole exciton features. The resulting 6 hyperspectral images corresponding to 6 different sample locations are available as attached Visualizations 1-6 in video format where time in the video maps to optical wavelength. A select spectral slice from one hyperspectral image corresponding to the Visualization 1 is shown here in Fig.\ \ref{fig:r0}. In addition, white-light images of the sample were recorded at both ~6 K and room temperature -- see Fig.~\ref{fig:r0}. Often in the analysis of hyperspectral images, it is useful to be able choose specific spatial locations and examine the corresponding spectrum. To demonstrate this capability, we selected 5 adjacent pixels, labeled in Fig.~\ref{fig:r0} and plot the corresponding FWM spectra in Fig.~\ref{fig:FWM}.

\section{Discussion}
\label{sec:discussion}

The heavy-hole and light-hole exciton spectral features (see Ref.~\cite{Cundiff2008} -- especially Fig.~2) are also present in the FWM hyperspectral images (see videos in Supplemental Material) -- however, they show up at different wavelengths/energies, particularly with respect to position (see Fig.~\ref{fig:FWM}). It is well known that the spectral locations of these features depend on confinement \cite{Cundiff2008}, temperature, electric field via the Stark effect (or in this case the quantum confined Stark Effect (QCSE) \cite{Miller1984}), and strain via direct changes to the band structure \cite{Wilmer2016}. Note that both GaAs and AlGaAs are non-centrosymmetric resulting in coupling between strain and electric field via the piezoelectric and inverse piezoelectric effects. Interestingly, neither Ref.~\cite{Wilmer2016} nor Ref.~\cite{Miller1984} mention the complimentary effect despite this coupling and despite each being capable of similar magnitude shifts of the exciton spectral features.

The source of the spatial dependence of the exciton FWM spectral features seen in Fig.~\ref{fig:FWM} and in the videos in the Supplemental Material can be assessed as follows. It is unlikely to be spatial temperature variation because there is no reason to suspect the sample is not uniformly cooled. Likewise, spatial confinement variation could explain the results but since the sample was grown using molecular beam epitaxy this is also unlikely. On the contrary, both a spatially varying QCSE resulting from/or a spatially varying strain could produce the range of shifts seen in the results. Visual inspection of the room-temperature and cryogenic white-light images in the videos in the Supplemental Material appears to reveal a temperature dependent straining of the quantum wells. The cryogenic images clearly show many spatial features that match with the hyperspectral images. The room-temperature images, however, show far fewer similarities - particularly only cracks within the sample. A temperature-dependent strain could be generated by a thermal expansion coefficient mismatch with the sapphire substrate. Without an independent control or measurement of either the strain or electric field, it is impossible to isolate the contributions from the strain-based and QCSE shifts.

As for the technique itself, the observed heavy-hole and light-hole features qualitatively agree with Ref.\ \cite{Cundiff2008} up to the shifts described above. Furthermore, the spatial features of the hyperspectral images qualitatively agree with the cryogenic white-light images. The primary limitation of this demonstration was the slow rate of data transfer between the DAQ board and the computer. Because of thermal drift of the sample, this limited the number of bursts that could be acquired for each pixel to less than 10 which ultimately limited the SNR. Furthermore, this low SNR and the short decay times of the excitons relative to the pulse repetition period made it unnecessary to process all of the data between bursts. We point out that these limitations are not fundamental to the technique and could be mitigated with several real-time correction schemes \cite{Roy2012,Ideguchi2013,Shen2018}. The SNR could also be improved by preventing the pumps beams from reaching the FWM-LO detector. This could be achieved, for example, by using the boxcars geometry \cite{Bristow2009}.

Even more spectral information could be obtained by performing MDCS \cite{Lomsadze2017b,Lomsadze2017c,Lomsadze2018b} instead of just spectrally resolved FWM. The most logical way to extend the technique presented here would be to utilize tri-comb spectroscopy \cite{Lomsadze2018}. Such a technique would generate enormous amounts of data and would almost certainly require real-time processing techniques. Lastly, we point out that this technique can simultaneously acquire linear data as well if desired.

\section{Conclusion}
\label{sec:conc}

The technique presented here is spectrally precise and potentially rapid. It is capable of generating near-diffraction-limited FWM hyperspectral images. Furthermore, it can be applied to any material in principle. To the best of our knowledge there is no other technique with all of these features.

\section*{Acknowledgments}
This work was supported by: National Science Foundation (NSF) (1256260); Intelligence Advanced Research Projects Activity (IARPA) (2016-16041300005).

This research is based on work supported by the Office of the Director of National Intelligence (ODNI), Intelligence Advanced Research Projects Activity (IARPA), via contract 2016-16041300005. The views and conclusions contained herein are those of the authors and should not be interpreted as necessarily representing the official policies or endorsements, either expressed or implied, of the ODNI, IARPA, or the U.S. government. The U.S. government is authorized to reproduce and distribute reprints for governmental purposes notwithstanding any copyright annotation thereon.

\bibliographystyle{titles}
\bibliography{References}

\end{document}